\documentclass[aps,prl,preprint,amsmath,amssymb,superscriptaddress]{revtex4}
\usepackage{graphicx}
\usepackage{dcolumn}
\usepackage{color}
\usepackage{bm}
\usepackage{amsmath,amssymb}
\usepackage[normalem]{ulem}

\newcommand{\vc}[1]{\mathbf{#1}}

\newcommand{\vk}{\vc{k}}

\newcommand{\unit}[1]{\text{#1}}

\newcommand{\BT}{\ensuremath{\mathrm{Bi}_2\mathrm{Te}_3}}
\newcommand{{\muB}}{\ensuremath{\mu_{\rm B}}}

\begin{document}



\title{Mesoscopic 
electron focusing in topological insulators }

\author{P.\,Sessi} 
	\email[corresponding author: ]{sessi@physik.uni-wuerzburg.de}
	\affiliation{Physikalisches Institut, Experimentelle Physik II, 
	Universit\"{a}t W\"{u}rzburg, Am Hubland, 97074 W\"{u}rzburg, Germany}
\author{P.\,R{\"u}{\ss}mann} 
	\affiliation{Peter Gr\"{u}nberg Institut and Institute for Advanced Simulation, 
	Forschungszentrum J\"ulich and JARA, 52425 J\"ulich, Germany}
\author{T.\,Bathon} 
	\affiliation{Physikalisches Institut, Experimentelle Physik II, 
	Universit\"{a}t W\"{u}rzburg, Am Hubland, 97074 W\"{u}rzburg, Germany}
\author{A.\,Barla} 
	\affiliation{Istituto di Struttura della Materia, Consiglio Nazionale delle Ricerche, 34149 Trieste, Italy}
\author{K.\,A.\,Kokh} 
	\affiliation{V.S. Sobolev Institute of Geology and Mineralogy, Siberian Branch, Russian Academy of Sciences,
	630090 Novosibirsk,  Russia}
	\affiliation{Novosibirsk State University, 630090 Novosibirsk,  Russia}
\author{O.\,E.\,Tereshchenko} 
	\affiliation{A.V. Rzanov Institute of Semiconductor Physics, Siberian Branch, 
	Russian Academy of Sciences, 630090 Novosibirsk,  Russia}
	\affiliation{Novosibirsk State University, 630090 Novosibirsk,  Russia}
\author{K.\,Fauth} 
	\affiliation{Physikalisches Institut, Experimentelle Physik II, 
	Universit\"{a}t W\"{u}rzburg, Am Hubland, 97074 W\"{u}rzburg, Germany}
	\affiliation{Wilhelm Conrad R{\"o}ntgen-Center for Complex Material Systems (RCCM), 
	Universit\"{a}t W\"{u}rzburg, Am Hubland, 97074 W\"{u}rzburg, Germany}
\author{S.\,K.\,Mahatha} 
	\affiliation{Istituto di Struttura della Materia, Consiglio Nazionale delle Ricerche, 34149 Trieste, Italy}
\author{M.\,A.\,Valbuena} 
	\affiliation{Catalan Institute of Nanoscience and Nanotechnology (ICN2), CSIC, 
	and The Barcelona Institute of Science and Technology, Campus UAB, Bellaterra, 08193 Barcelona, Spain}
\author{S.\,Godey} 
	\affiliation{Catalan Institute of Nanoscience and Nanotechnology (ICN2), CSIC, 
	and The Barcelona Institute of Science and Technology, Campus UAB, Bellaterra, 08193 Barcelona, Spain}
\author{A.\,Mugarza} 
	\affiliation{Catalan Institute of Nanoscience and Nanotechnology (ICN2), CSIC, 
	and The Barcelona Institute of Science and Technology, Campus UAB, Bellaterra, 08193 Barcelona, Spain}
	\affiliation{ICREA--Institucio Catalana de Recerca i Estudis Avancats, Lluis Companys 23, 08010 Barcelona, Spain}
\author{P.\,Gargiani} 
	\affiliation{ALBA Synchrotron Light Source, 08290 Cerdanyola del Vall�s, Barcelona, Spain}
\author{M.\,Valvidares} 
	\affiliation{ALBA Synchrotron Light Source, 08290 Cerdanyola del Vall�s, Barcelona, Spain}		
\author{N.\,H.\,Long} 
	\affiliation{Peter Gr\"{u}nberg Institut and Institute for Advanced Simulation, 
	Forschungszentrum J\"ulich and JARA, 52425 J\"ulich, Germany}
\author{C.\,Carbone} 
	\affiliation{Istituto di Struttura della Materia, Consiglio Nazionale delle Ricerche, 34149 Trieste, Italy}
\author{P.\,Mavropoulos} 
	\affiliation{Peter Gr\"{u}nberg Institut and Institute for Advanced Simulation, 
	Forschungszentrum J\"ulich and JARA, 52425 J\"ulich, Germany}
\author{S.\,Bl{\"u}gel} 
	\affiliation{Peter Gr\"{u}nberg Institut and Institute for Advanced Simulation, 
	Forschungszentrum J\"ulich and JARA, 52425 J\"ulich, Germany}
\author{M.\,Bode}
	\affiliation{Physikalisches Institut, Experimentelle Physik II, 
	Universit\"{a}t W\"{u}rzburg, Am Hubland, 97074 W\"{u}rzburg, Germany}
	\affiliation{Wilhelm Conrad R{\"o}ntgen-Center for Complex Material Systems (RCCM), 
	Universit\"{a}t W\"{u}rzburg, Am Hubland, 97074 W\"{u}rzburg, Germany}



\date{\today}
\begin{abstract}
\vspace{1cm}
\bf The particle-wave duality sets a fundamental correspondence 
	between optics and quantum mechanics. 
	Within this framework, the propagation of quasiparticles 
	can give rise to superposition phenomena which, 
	like for electromagnetic waves, can be described by the Huygens principle. 
	However, the utilization of this principle by means of propagation and manipulation 
	of quantum information is limited by the required coherence in time and space. 
	Here we show that in topological insulators, which in their pristine form are characterized 
	by opposite propagation directions for the two quasiparticles spin channels,  
	mesoscopic focusing of coherent charge density oscillations can be obtained 
	at large nested segments of constant-energy contours by magnetic surface doping. 
	Our findings provide evidence of strongly anisotropic Dirac fermion-mediated interactions. 
	Even more remarkably, the validity of our findings goes beyond topological insulators 
	but applies for systems with spin-orbit--lifted degeneracy in general.  
	It demonstrates how spin information can be transmitted over long distances, 
	allowing the design of experiments and devices based on coherent quantum effects 
	in this fascinating class of materials.	
\noindent
\end{abstract}


\maketitle
\newpage
Coherence is a general property of waves as it describes the capability of keeping 
a well-defined phase relation while propagating in space and time. 
Because of the particle-wave duality, which lays at the very foundations of quantum mechanics, 
the same concept can also be applied to quasiparticles in solids. 
Quantum coherence is of fundamental importance since it sets the limits 
up to which information can be transmitted and processed with high fidelity.  
With the invention of the scanning tunneling microscope it became possible 
to visualize coherent phenomena in real space by imaging the standing wave pattern 
produced by scattering events around individual atomic-scale defects \cite{CLE1993}. 
In analogy with electromagnetic waves these results can be interpreted within the Huygens principle. 
It describes the interference pattern which results from the superposition of waves 
propagating along all different paths and can be theoretically elegantly expressed 
by using the quantum-mechanical propagator.  

The further development of atomic-scale manipulation techniques 
allowed to engineer these properties at the atomic scale.    
This capability was used for the creation of exotic effects such as quantum mirages \cite{MLE2000}, 
for the extraction of the phase of electron wave functions \cite{MMF2008}, 
and for visualizing the indirect coupling mechanisms 
mediated by conduction electrons such as the RKKY interaction \cite{ZWL2010}.
More recently, quantum interference imaging allowed to analyze 
how propagating waves in solids are influenced by the periodic potential of the crystal lattice. 
In particular, it has been shown that the propagation of quasiparticle waves can become anisotropic 
when the shape of a constant-energy cut (CEC) deviates from an isotropic contour.  
In analogy to optics, focussing and defocussing 
lead to an enhanced intensity along certain crystallographic directions 
and to partial or even complete suppression along others, respectively \cite{WWL2009}. 
However, despite its relevance in several areas of modern condensed matter, 
spintronics and quantum computation being the two most remarkable examples, 
the role played by the spin degree of freedom has not yet been explored. 

Within this framework, the recently discovered topological insulators (TIs)
represent a promising class of materials. 
TIs are insulating in the bulk but conducting on the surface 
where they host linearly dispersing massless fermions \cite{HK2010}. 
The strong spin-orbit coupling locks the spin to the momentum, 
thereby suppressing backscattering \cite{RSP2009,ZCC2009},
and results in spin currents which are intrinsically tied to charge currents \cite{KWB2007}. 
The overwhelming majority of studies discussed electronic states in the vicinity of the Dirac point 
where the isotropic linear band dispersion relation results in circular CEC. 
However, both photoemission and theoretical studies showed that iso-energy cuts 
progressively evolve from an almost circular (and convex) shape very close to the Dirac energy 
to a more concave, snowflake-like shape at higher energies \cite{CAC2009,F2009}. 
Once this transition takes place, large parallel sections face each other, 
a scenario supporting good nesting vectors which can strongly enhance 
the susceptibility of the system to external perturbations. 

In pristine TIs, an active role of these vectors is strongly suppressed 
by time reversal symmetry \cite{RSP2009,ZCC2009}. 
Here, we demonstrate that once this protection is lifted 
by the introduction of magnetic surface dopants, 
the nesting leads to strongly focused interference patterns in the charge density, 
resulting in coherent quantum oscillations which can be observed 
by quasiparticle interference (QPI) imaging with the scanning tunneling microscope (STM)
over distances of tens of nanometers without any significant intensity loss. 
Theoretical calculations rationalize these findings in terms of the combined action 
of the iso-energy contour shape and the local magnetic moments present on the surface, 
thereby providing guidelines to control this effect. 
In particular, for a long coherence length two conditions need to be fulfilled: 
(i) the Fermi energy contour must exhibit large nearly parallel segments (nesting) 
with pairs of initial and final $k$-points which can be connected by the same scattering vector, 
and (ii) the magnetic dopants must couple to generate a high-spin state 
with a magnetic moment well beyond a single atom. 
As revealed by x-ray magnetic circular dichroism measurements (XMCD) 
this is fulfilled already at very dilute Mn concentrations on Bi$_2$Te$_3$. 
Our observations provide evidence that the emergence of superparamagnetic order 
on magnetically doped TIs with nested constant-energy contours 
can trigger Dirac fermion--mediated highly anisotropic indirect interactions. 
These results suggest that---by appropriate band engineering---spin-dependent 
quantum coherent transport can be achieved over mesoscopic distances 
and pave the way to design novel device concepts that rely on 
quantum coherent effects in this fascinating class of materials.    

\begin{figure*}[h]   
\includegraphics[width=\textwidth]{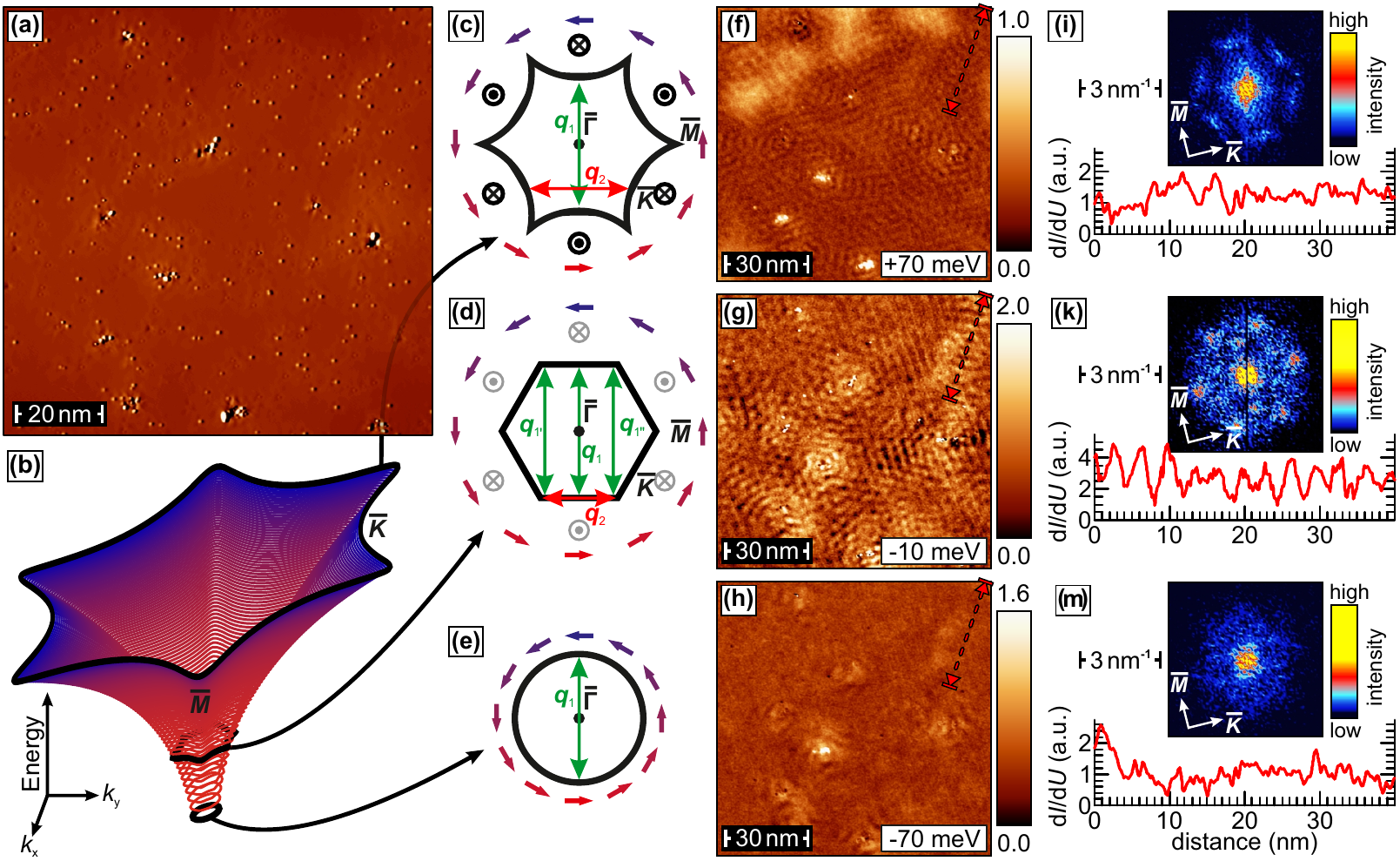}%
\caption{
	(a) Topographic STM image of 0.01\,ML Mn on Bi$_2$Te$_3$ ($I = 35$\,pA; $U = 150$\,mV).
	(b)~Schematic dispersion of the topological states above the Dirac point. (c)--(e)
	Constant-energy contours with in-plane/out-of-plane spin polarization 
	at selected energies. 
	Red and green arrows indicate the preferred scattering vectors $\mathbf q_1$ and $\mathbf q_2$ 
	along $\overline{\Gamma K}$ and $\overline{\Gamma M}$ directions, respectively. 
	Corresponding quasiparticle interference maps measured at energies 
	that mark the transition from convex to concave constant energy cuts are shown 
	in (f) $E - E_{\rm F} = +70$\,meV, (g) $-10$\,meV, and (h) $-70$\,meV. 
	Line sections measured between the arrows along the dashed red line are plotted in panels (i)--(m).  
	In addition, they show the corresponding Fourier-transformed $\mathrm{d}I/\mathrm{d}U$ maps (top, raw data).   
}
\label{QPI}
\end{figure*}  
Figure\,\ref{QPI}(a) shows the constant-current STM image of a Bi$_2$Te$_3$ crystal 
which was surface-doped with Mn by e-beam evaporation onto the cold sample surface. 
The Mn coverage amounts to approximately 0.01 monolayer (ML) 
and gives rise to a rigid negative energy shift 
of the band structure of approximately 100\,meV, resulting in an n-doped surface with the Dirac point 
energetically located about 250\,meV below the Fermi level~$E_{\rm F}$ \cite{SRB2014}. 
Note, that this leads to a Fermi level positioned inside the bulk band gap, 
making the Fermi sea a truly two-dimensional system (see Supplementary Information).   
The dispersion above the Dirac point is schematically represented in Fig.\,\ref{QPI}(b). 
Selected CECs relevant for the following discussion 
are sketched in Fig.\,\ref{QPI}(c)-(e). 
Just above the Dirac point [Fig.\,\ref{QPI}(e)] the CEC is circular (black line)
and the spin is perpendicularly locked to the momentum, 
thereby leading to a helical spin structure indicated by colorized arrows. 
As we move further away from the Dirac point the warping increases.  
This first leads to a hexagonal [Fig.\,\ref{QPI}(d)] and eventually 
to a snowflake-like shape of the CEC [Fig.\,\ref{QPI}(c)]. 
This deformation goes along with the development 
of an alternating out-of-plane component of the spin polarization 
along $\overline{\Gamma K}$ directions of the surface Brillouin zone \cite{F2009}, 
as indicated by symbols ($\otimes,\odot$) in Fig.\,\ref{QPI}(c) and (d).  

Panels (f)--(h) of Fig.\,\ref{QPI} show quasiparticle interference ($\mathrm{d}I/\mathrm{d}U$) maps 
measured at three energies, $E - E_{\rm F} = +70$\,meV, $-10$\,meV, and $-70$\,meV, respectively.   
These energies are chosen such that they mark the transition from convex to concave CEC, 
as consistently shown by theoretical calculations \cite{F2009} 
and photoemission experiments \cite{CAC2009}. 
As indicated by green and red arrows in \mbox{Fig.\,\ref{QPI}(c)-(e)}, 
the scattering vectors $\mathbf q_1$ and $\mathbf q_2$ specific for each energy can be obtained 
by using the stationary phase approximation \cite{LQZ2012}. 
At $E - E_{\rm F} = -70$\,meV [Fig.\,\ref{QPI}(e) and (h)] the CEC is almost circular. 
Since reasonably nested parts 
essentially contain only one scattering vector, $\mathbf q_1$,  
only very weak QPI modulations are observed in Fig.\,\ref{QPI}(h).  
Indeed, the line section taken along the red line in Fig.\,\ref{QPI}(h)
as well as the Fourier-transformed \mbox{(FT-)QPI} map, both displayed in Fig.\,\ref{QPI}(m), 
confirm the very low intensity of scattering events at this bias voltage.   
Note that this backscattering vector $\mathbf q_1$ is completely prohibited 
in the pristine material since reversing the wave vector would require a spin-flip, 
a mechanism forbidden as long as time-reversal symmetry is preserved. 

In contrast, very strong scattering is observed in the QPI map 
recorded at \mbox{$E - E_{\rm F} = -10$\,meV} [Fig.\,\ref{QPI}(g)] 
where the CEC becomes hexagonal. 
Intense modulations of the $\mathrm{d}I/\mathrm{d}U$ signal 
can be recognized and quantified in the line profile [see Fig.\,\ref{QPI}(k)].  
The FT-QPI map reveals the simultaneous existence of two scattering vectors 
which---as a result of the hexagonal surface symmetry---both appear in six equivalent directions, 
here represented by six (very weak) inner spots in $\overline{\Gamma M}$ directions 
and six (stronger) outer spots which can be found along $\overline{\Gamma K}$ directions.  

The scattering channel along $\overline{\Gamma M}$ directions ($\mathbf q_2$)
is routinely found on TI materials \cite{ZCC2009,SOB2013}.  Its appearance is related to the strength of the warping term which allows to effectively scatter between next-nearest neighbors segments of a CEC because of their parallel spin polarization.
This channel is rather weak at the energy discussed here, $E - E_{\rm F} = -10$\,meV, 
since nesting along $\mathbf q_2$ is still rather poor [see scheme in Fig.\,\ref{QPI}(d)].  
In contrast, high-intensity spots are visible along the $\overline{\Gamma K}$ directions, which cannot be found on pristine TI surfaces.  
Their appearance implies an active role of the time-reversal symmetry breaking perturbations, i.e. the Mn adatoms, which, following theoretical predictions \cite{LLX2009}, we recently suggested to be magnetically coupled  by the Dirac fermions present on the TI surface \cite{SRB2014} (see XMCD discussion below).  

Further raising the energy to $E - E_{\rm F} = +70$\,meV  
leads to a weaker modulation of the $\mathrm{d}I/\mathrm{d}U$ signal 
detected in QPI images (see scale bars of Fig.\,\ref{QPI}). This finding is confirmed by the FT-QPI map shown in the top panel of Fig.\,\ref{QPI}(i) 
which only reveals six weak spots along the $\overline{\Gamma M}$ direction corresponding to $\mathbf q_2$.

Interestingly, closer inspection of the real space images reveals that, 
once backscattering channels are opened, coherent waves propagate 
without any significant intensity loss over distances larger than 30\,nm, 
as can clearly been seen in the line section of Fig.\,\ref{QPI}(k). 
Furthermore, although point-shape scattering centers should result 
in spherical waves emanating in all radial directions, 
our experimental data presented in Fig.\,\ref{QPI}(g) indicate 
that the charge density oscillations remain highly focused 
over mesoscopic distances well beyond the atomic scale. 
These observations provide compelling evidence that Dirac fermion--mediated interactions 
in TIs are highly anisotropic, where some crystallographic directions are preferred over others, 
as predicted theoretically \cite{BB2010}. 
Even more remarkably they demonstrate that, by appropriate band engineering, spin coherence 
can be achieved over mesoscopic distances.  

A priori, however, the exact mechanisms that lead to the emergence of highly focused and anisotropic QPI patterns are not evident. 
Good nesting is known to be necessary to trigger the focusing effect
\cite{WWL2009,Lounis2011}.  In the present case, this condition
is fulfilled by large portions of the FS especially for the hexagonal case. This configuration supports
additional scattering vectors with $\mathbf{v}_{\mathbf{k}}\approx
-\mathbf{v}_{\mathbf{k}'}$ ($\mathbf{v}_{\mathbf{k}}$ is the group velocity), such as $\mathbf q_1^{\prime}$ or $\mathbf q_1^{\prime\prime}$, 
which---in contrast to $\mathbf q_1$---are not strictly forbidden by time-reversal symmetry 
but strongly suppressed by the almost opposite spin projection of initial and final states. Their absence in the pristine case proves the important role played by the magnetic perturbations which needs thus to be carefully analyzed.
Several different scenarios can be invoked: 
(i)~the existence of time-reversal symmetry breaking perturbations is sufficient;
(ii)~both conditions, FS nesting \emph{and}  the presence of magnetic scatterers, 
are required;  
(iii)~the Mn adatoms effectively polarize the topological-state spins. 

As we will show by means of {\it ab-initio} as well as by model-Hamiltonian simulations, only
scenario (ii) correctly explain our experimental findings.

The quantity that we examine is the extended joint density of states (exJDOS), 
an extension of the widely used joint density of states approach \cite{Eich2014,RSP2009} 
that constitutes an approximation to Fourier-transformed $\mathrm{d}I/\mathrm{d}U$ images. 
We define the exJDOS as the convolution
\begin{equation}
\label{eq:exJDOS}
{\rm exJDOS}(\vc{q};E) = 
\int  A^{\rm surf}_{\vc{k}}(E) M_{\vc{k},\vc{k}+\vc{q}}(E) A^{\rm surf}_{\vc{k}+\vc{q}}(E) \mathrm{d}^2\vc{k},
\end{equation} 
at energy $E$, involving the spectral density $A^{\rm surf}_{\vk}(E)$ 
integrated in the spatial region between tip and sample surface, 
and the matrix element $M_{\vk\vk'}(E) = P_{\vk\vk'}(E)\, \gamma_{\vk\vk'}(E)$. 
Here, $P_{\vk\vk'}(E)$ is the scattering rate (including the suppression of time-reverse scattering) 
and $\gamma_{\vk\vk'}(E)=1-\cos(\vc{v}_{\vk},\vc{v}_{\vk'})$ 
accounts for the fact that the STM probes standing waves, 
i.e., states with opposite group velocities are favored, in the spirit of the stationary phase approximation. 
The quantities $A^{\rm surf}_{\vk}(E)$ and $\gamma_{\vk\vk'}(E)$ are derived from the band structure, 
while $P_{\vk\vk'}(E)$ is calculated by means of the Golden Rule 
from the $T$-matrix of the impurity obtained by a calculation of the impurity Green function.

Our density-functional calculations are based on the local density approximation \cite{Vosko1980}. 
We employ the Korringa-Kohn-Rostoker Green-function (KKR) method 
for the calculation of the electronic structure and scattering properties 
($T$-matrix and surface state scattering rate) of the impurity. 
The \BT\ surface is modeled by a film of 6 quintuple layers. 
\begin{figure}
	\begin{minipage}[t]{0.6\textwidth}
	\includegraphics[width=0.99\columnwidth]{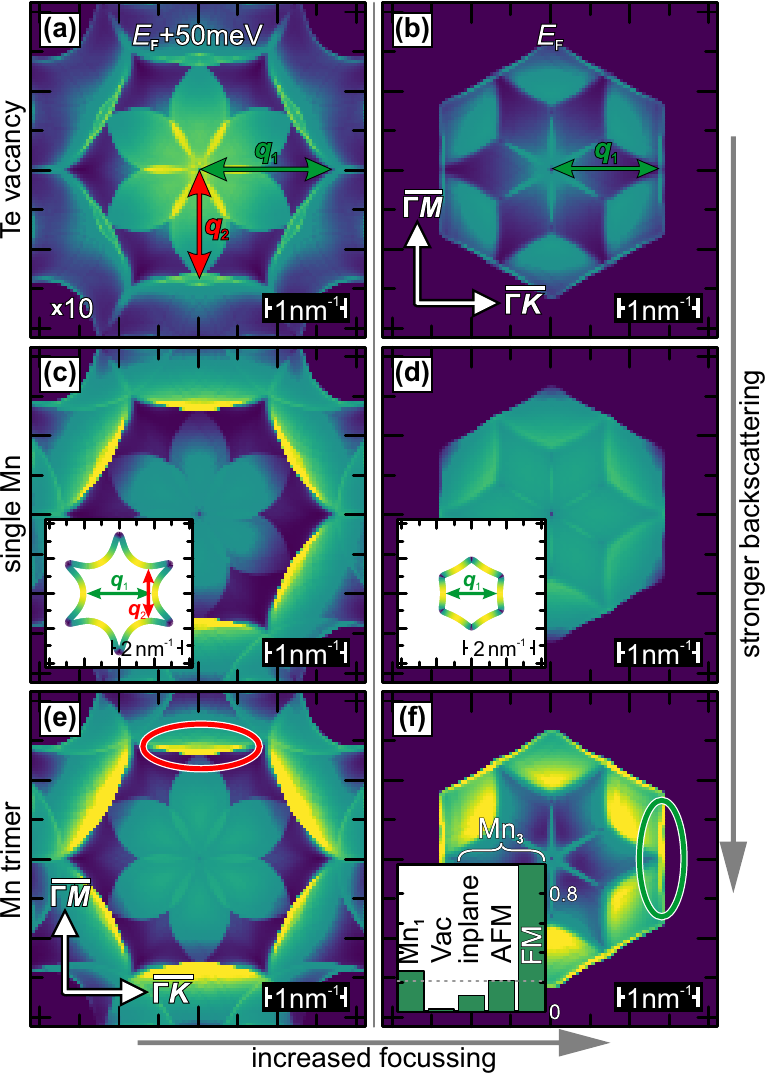}
	\end{minipage}
	\hfill
	\begin{minipage}[b]{0.39\textwidth}
	\caption{
	Simulated exJDOS images for energies $E = E_F + 50 \unit{meV}$ (a, c, e) and $E = E_F$ (b, d, f) 
	as generated by a Te vacancy (top row), a single Mn atom at a Te vacancy site (middle), 
	and a Mn trimer (bottom). 
	With the exception of (a) (contrast 10-fold magnified) the color scale is the same in all panels.
	Red ($\overline{\Gamma M}$ direction) and green ($\overline{\Gamma K}$) ellipses 
	highlight exJDOS features associated to conventional scattering and backscattering, respectively. 
	Insets in (c) and (d) show iso-energy contours for the different energies 
	color-coded by the states' surface localization (yellow: $>80\%$ in first QL, dark blue: $\sim60\%$ in first QL). 
	In the inset in (f) the relative magnitude of the backscattering signal 
	is compared for 5 different impurity and spin configurations.}
\label{fig:DFT} \end{minipage}    
\end{figure}
Fig.~\ref{fig:DFT} summarizes a number of KKR-based numerical experiments 
showing the exJDOS at different energies 
[corresponding to warped (left) or hexagonal (right) Fermi surface] 
and induced by different defects/impurities [top: non-magnetic Te vacancy; 
middle: single magnetic Mn impurity (4.4\,{\muB}) substituting Te; 
bottom: magnetic Mn trimer filling three Te vacancies in a ferromagnetic configuration (13.1\,{\muB})]. 
If backscattering (green arrows) is present, a situation observed experimentally in Fig.\,\ref{QPI}(g), 
we expect a high exJDOS$(\vc{q};E)$ intensity along $\overline{\Gamma K}$ directions.  
Evidently this is only the case for Fig.\,\ref{fig:DFT}(f), i.e., when the nesting condition is fulfilled 
{\em and} when a very strong impurity magnetic moment (13.1\,{\muB}\ of the Mn trimer) is present. 
Such a high magnetic moment can only be reached 
for a Mn trimer with strong ferromagnetic correlations.  
Other configurations (not shown) with the atomic moments within the Mn trimer 
oriented along different directions, such as $\uparrow \uparrow \downarrow$ 
(resulting in a magnetic moment of 4.4\,{\muB}) or three in-plane--oriented spins 
in a $120^\circ$ configuration (vanishing net moment; 0\,{\muB}), 
also show a negligeble backscattering intensity. 
The relative intensity of the Te vacancy (vac.), the single Mn atom (Mn$_1$), 
and Mn trimers with different spin configurations 
[in-plane $120^\circ$, antiferromagnetic (AFM), and ferromagnetic (FM)]
is depicted in the histogram shown in the inset of Fig.~\ref{fig:DFT}(f). 
Indeed, our {\it ab initio} density functional theory calculations performed 
for a scenario where three Mn atoms are filling three Te vacancies show that the ferromagnetic state 
has the lowest total energy as compared to the other configurations discussed above. 
We speculate that this may be the result of indirect RKKY-type coupling 
mediated by the Dirac fermions which exhibit a Fermi wavelength of approximately 7\,nm, 
i.e.\ well above the average Mn--Mn spacing of about 3\,nm, making the interaction always ferromagnetic as predicted in Ref.\cite{LLX2009}.

In parallel we performed model-calculations based on the Hamiltonian of Lee \textit{et al.} \cite{Lee2009}. 
As described in the supplementary information, the model was extended 
by exchange-interaction terms to represent (i) the scattering by Mn moments, 
as well as (ii) the possibility of a uniform magnetization of the surface state 
caused by ferromagnetic coupling among the Mn defects. 
These model calculations, in agreement with the {\em ab-initio} results, provide further evidence 
that the experimentally observed standing wave patterns arise 
from the combined action of ferromagnetically coupled Mn atoms and the hexagonally shaped CEC supporting the focusing effect. 

\begin{figure}
\centering
\includegraphics[width=\columnwidth]{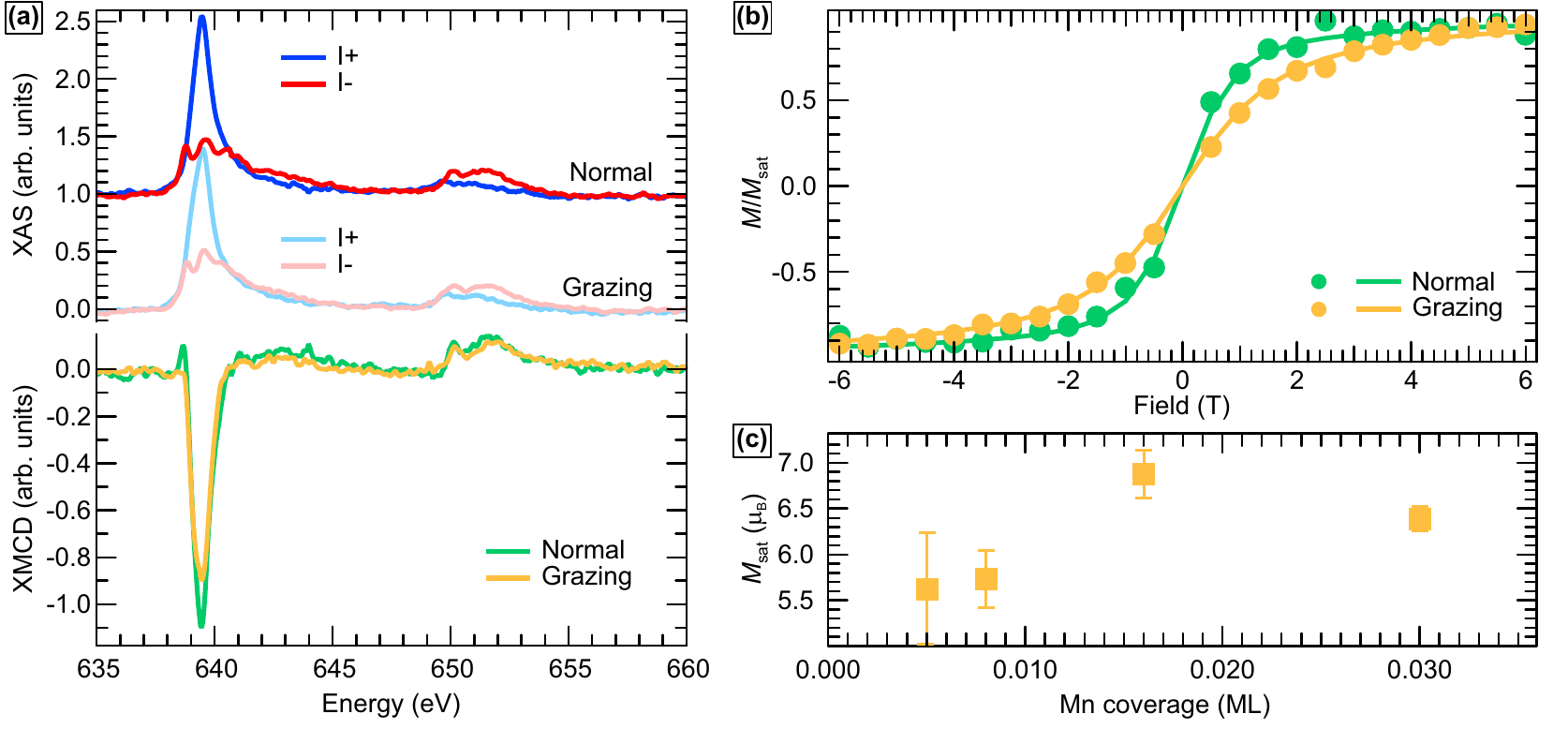}
\caption[]{(a) Mn $L_{2,3}$-edges x-ray absorption spectra (XAS; top panel)
and x-ray magnetic circular dichroism (XMCD; bottom panel) signal
of 0.016\,ML Mn on Bi$_2$Te$_3$, obtained at normal
and grazing incidence, with $B = 6$\,T and $T = 2.5$\,K.
(b) Corresponding magnetization cycles, recorded at the XMCD maximum at the Mn $L_3$-edge.
(c) Mn-coverage dependence of the saturation magnetization
extracted from the recorded magnetization cycles.}
\label{fig:XMCD}
\end{figure}

\bigskip
Achieving such a high magnetic moment in a Mn-doped system
is possible only if Mn atoms couple ferromagnetically since both 
bulk Mn \cite{LLA1994} and Mn nanostructures \cite{SGG2009},
are known to exhibit an antiferromagnetic ground-state.
In order to directly determine the magnetic moment and configuration of Mn on Bi$_2$Te$_3$,
we have performed XMCD measurements
at the BOREAS beamline of the Alba synchrotron facility.
By measuring the photo-induced sample drain current, x-ray absorption spectra (XAS) were recorded
in the total electron yield (TEY) mode with left ($I^-$) and right ($I^+$) circularly polarized photons
and in the presence of an external magnetic field.
Fig.\,\ref{fig:XMCD}(a) (top panel) reports the XAS recorded at the Mn $L_{2,3}$ edges
with normal and grazing x-rays incidence, at a Mn coverage of about 0.016\,ML.
The data were taken at a temperature $T = 2.5$\,K
and in a magnetic field of $6$\,T applied along the photon beam direction.
The spectra are characteristic of Mn atoms in a configuration close to $d^5$,
as previously observed for Mn doped into the bulk of Bi$_2$Te$_3$\,\cite{Watson2013}.

The XMCD, calculated as $I^- - I^+$ and shown in the bottom panel of Fig.\,\ref{fig:XMCD}(a),
highlights the considerable magnetic polarization of the Mn surface dopants.
Fig.\,\ref{fig:XMCD}(b) displays the magnetization cycles, 
recorded on the same sample at $T = 2.5$\,K, at normal (green line) and grazing incidence (yellow line), 
by following the XMCD magnitude at the Mn $L_3$-edge (full circles) as a function of the applied magnetic field.
Both, the larger slope in $M(H)$ at small fields and the saturation at large fields 
in normal incidence geometry indicate an out-of-plane magnetic anisotropy for Mn on Bi$_2$Te$_3$.
The saturation magnetization (in Bohr magnetons per Mn atom) can be evaluated 
by applying the XMCD sum rules \cite{Thole1992, Carra1993} to the data in Fig.\,\ref{fig:XMCD}(a).
Independent information on the magnitude of the fluctuating total moments 
is contained in the shape of the magnetization isotherms, strongly determining their slopes near $\mu_0 H = 0$.
An analysis in terms of classical Langevin paramagnetism 
augmented with a uniaxial magnetic anisotropy term \cite{Gambardella2003},
yields an effective value of the (\textit{total}) saturation moment $M_{\mathrm{sat}}$
and the associated magnetic anisotropy energy.

Fig.\,\ref{fig:XMCD}(c) shows the coverage dependence of $M_{\mathrm{sat}}$ 
in the range 0.005 -- 0.03\,ML Mn, resulting from the Langevin fits.
At the lowest coverage of 0.005\,ML we obtain an effective moment near 5\,{\muB}, 
coinciding with the saturation moment of Mn atoms with $d^5$ configuration.
In agreement with expectation the adatoms can be considered as magnetically independent in this limit.
Already at a coverage of 0.008\,ML, however, a significant enhancement 
of the effective magnetic moment becomes evident.  
Further increasing the coverage to 0.016\,ML leads to an effective moment of almost 7\,{\muB}.  
Such a large value cannot be ascribed to single Mn atoms.  
Instead, it implies the existence of ferromagnetic interactions among neighboring Mn atoms,
yielding a significant fraction of magnetic units constituted by more than one Mn atom.

We envision that exchange coupling between Mn adatoms may be mediated
by the surface electron gas of Bi$_2$Te$_3$.
We speculate that---although ferromagnetic correlations are present---thermal fluctuations
and disorder are too strong to establish stable ferromagnetic order.
Nevertheless, assemblies of Mn atoms with small enough Mn--Mn separation  
will exhibit sufficiently strong magnetic interactions to couple their individual moments to a macro-spin,
resulting in the enhanced susceptibility characteristic for superparamagnets
which is experimentally observed in Fig\,\ref{fig:XMCD}(b).  
Since XMCD spatially averages over a macroscopic sample area, 
a value of 7\,{\muB} implies the existence of assemblies with considerably larger magnetic moments.  
This result is consistent with the theoretical finding that units composed
of at least three ferromagnetically interacting Mn atoms are required
to activate the scattering channel along the $\overline{\Gamma K}$ direction. 

Our findings provide evidence that, through electron focusing, 
quantum coherent information can be transferred in topological insulators 
over distances of more than 30 nm, making it compatible with device dimensions 
and thereby paving the way to design experiments and devices 
based on spin quantum coherent phenomena in this fascinating class of materials. 
More generally, they provide evidence that, through appropriate band engineering 
in materials with spin-split states, many interesting phenomena may appear, 
with important implications for spintronic and quantum computation.

\section*{Acknowledgments}
P.S., T.B., and M.B.\ (BO 1468/21-1) and P.R., P.M., and S.B.\ (MA 4637/3-1)
acknowledge financial support through SPP 1666 from the Deutsche Forschungsgemeinschaft. 
The work has been partially supported by the Italian Government 
(MIUR Progetto Premiale ``Materiali e disposivi magnetici e superconduttivi per sensoristica e ICT'').
M.A.V., S.G., and A.M.\ acknowledge support from the 
Ministerio de Ciencia e Innovacion (Grant No. MAT2013-46593-C6-5-P) 
and the Severo Ochoa Program (MINECO, grant SEV-2013-0295). 
P.R., N.H.L., P.M. and S.B. acknowledge financial support from the VITI project of the 
Helmholtz Association as well as computational support from the JARA-HPC 
Supercomputing Centre at the RWTH Aachen University.


\bibliography{Focusing_Lit_02}

\end{document}